\documentclass[printer]{aa}  
\usepackage{graphicx}
\usepackage{txfonts}
\usepackage{natbib}
\begin{document}
\title{TYC~2675-663-1: A newly discovered W UMa system in an active state} 
\subtitle{}
\titlerunning{The new binary system TYC~2675-663-1}
\authorrunning{M.~D.~Caballero-Garcia et~al.}
\author{M. D. Caballero-Garc\'{\i}a \inst{1} \and G. Torres \inst{2} \and I. Ribas \inst{3} \and D. R\'{i}squez \inst{4} \and B. Montesinos \inst{5}
           \and J. M. Mas-Hesse \inst{5}
          }
   \offprints{M. D. Caballero-Garc\'{\i}a; \email{mcaballe@ast.cam.ac.uk} }

         \institute{University of Cambridge, Institute of Astronomy, Cambridge CB3 0HA, UK
         \and
              Harvard-Smithsonian Center for Astrophysics, 60 Garden St., Cambridge, MA 02138, USA
         \and
              IEEC (Institut d'Estudis Espacials de Catalunya), Edif. Nexus-104,
              Gran Capit\`a 2-4, E-08034 Barcelona, Spain
         \and
              Leiden Observatory, P. O. Box 9513, NL-2300 RA Leiden, The Netherlands
         \and
              CAB--LAEX (CSIC--INTA), ESAC Campus, P. O. Box 78, 28691 Villanueva de la Ca\~nada, Madrid, Spain
             }

\date{Received October 15, 2009; accepted January 25, 2010}

\abstract
{}
{The recently discovered eclipsing binary system TYC~2675-663-1 is a
    X-ray source, and shows properties in the optical that are similar
    to the W~UMa systems, but are somewhat unusual compared to what is
    seen in other contact binary systems. The goal of this work is to
    characterize its properties and investigate its nature by means of
    detailed photometric and spectroscopic observations.}
{We have performed extensive $V$-band photometric measurements with
the INTEGRAL satellite along with ground-based multi-band photometric
observations, as well as high-resolution spectroscopic monitoring from
which we have measured the radial velocities of the components. These
data have been analysed to determine the stellar properties, including
the absolute masses and radii. Additional low-resolution spectroscopy
was obtained to investigate spectral features.}
{From the measured eclipse timings we determine an orbital period for
  the binary of $P=0.4223576 \pm 0.0000009$ days. The light-curve and
  spectroscopic analyses reveal the observations to be well
  represented by a model of an overcontact system composed of
  main-sequence F5 and G7 stars (temperature difference of nearly
  1000~K), with the possible presence of a third star.
  Low-resolution optical spectroscopy reveals a complex H$_\alpha$
  emission, and other features that are not yet understood.  The
  unusually large mass ratio of $q=0.81 \pm 0.05$ places it in the
  rare ``H'' (high mass ratio) subclass of the W~UMa systems, which
  are presumably on their way to coalescence.}
{}

\maketitle

%

\section{Introduction}

The star TYC~2675-663-1 \citep[Tycho-2 catalog
designation;][]{hog00}, with coordinates $\alpha = 20^{\rm h} 09^{\rm
m} 11.2^{\rm s}$, ${\delta}=+32^{\circ} 33\arcmin 53\arcsec$ (J2000),
was observed during 2002--2005 in the Johnson $V$ filter with the
Optical Monitoring Camera (OMC) on board the INTEGRAL satellite
\citep{mas-hesse03},
as a part of a serendipitous program to monitor optical counterparts
of ROSAT sources as potential variable objects. By cross-correlating
the list of variable sources detected with the OMC during the first
few months of operation with the ROSAT catalogs, five were found
that were potential optical counterparts of ${\rm X}$-ray sources
(i.e., that were inside the ROSAT 3$\sigma$ position error radius;
\citealt{caballero04, caballero06}). Three of them showed optical
variability with modulations characteristic of binary systems
(periodic variations, with regular periods and amplitudes larger than
0.1\,mag). TYC~2675-663-1 showed the most striking variability
pattern, with very irregular and variable behaviour.

The source, which has received the designation IOMC~2675000078 in the
{\it OMC Input Catalogue} \citep{domingo03} displays a colour index of
$B\!-\!V \approx 0.7$ \citep{hog00}, corresponding roughly to spectral
type G, and shows variability typical of a close eclipsing binary
system. Little further information is available for this star, aside
from entries in various astrometric catalogs, so we undertook
photometric and spectroscopic observations in order to ascertain the
nature of the object and identify the origin of its X-ray emission.

In this paper we present a detailed study of TYC~2675-663-1. The
photometric and spectroscopic observations are presented in
Section~\ref{observations}, with a description of the data reduction
procedures.  Section \ref{analysis} contains the analysis of these
observations, including the determination of the ephemeris, the
stellar parameters and distance to the system, and a discussion of the
peculiar H$_{\alpha}$ emission we detect. Finally, in
Section~\ref{discuss} we discuss the results and the nature of the
system.


\section{Observations and data reduction}  \label{observations}

\subsection{Photometry} \label{photometry}

\subsubsection{INTEGRAL/OMC photometry} \label{omc_phot}

The Optical Monitoring Camera (OMC) is a 50 mm aperture refractor
telescope, co-aligned with the high-energy instruments on board the
ESA INTEGRAL gamma-ray observatory \citep{mas-hesse03}. The OMC
provides continuous monitoring of up to 100 sources in the Johnson $V$
band over its $5^\circ \times 5^\circ$ field of view. Telemetry
constraints do not permit downloading of the entire OMC image.  For
this reason, windows were selected around the proposed
X-ray/$\gamma$-ray targets as well as other targets of interest in the
same field of view.  The position of these windows is computed
automatically, based on the sources compiled in the {\it OMC Input
Catalogue} \citep{domingo03}, which contains around 500\,000 targets
selected for being potentially variable in the optical. Only
sub-windows of the CCD containing those objects, with a size of $11
\times 11$ pixels ($3.2\arcmin \times 3.2\arcmin$), are transmitted to the
ground. TYC~2675-663-1 is located at only 14$\arcsec$ from the ROSAT
source 1RXS J200912.0+323344 (contained in the {\it ROSAT All-Sky
Survey Bright Source Catalogue}, \cite{voges99}, with a catalog
1$\sigma$ position error of 8\arcsec), and this allowed us to monitor
it as a potential optical counterpart.

\begin{figure*}
   \centering
   \includegraphics[bb=35 267 560 575,width=10cm,clip]{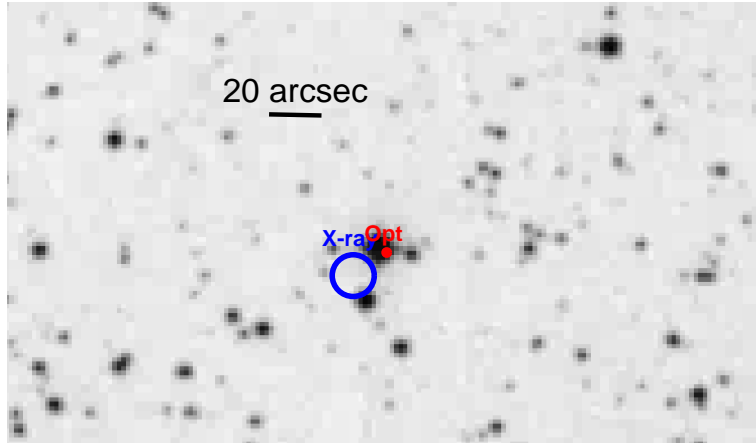}
      \caption{Field of view of the INTEGRAL/OMC target TYC~2675-663-1
              (red point) and the ROSAT source 1RXS J200912.0+323344
              (blue circle) at its nominal catalog position.}
         \label{field}
   \end{figure*}

\begin{figure*}
\begin{center}
\includegraphics[width=8cm]{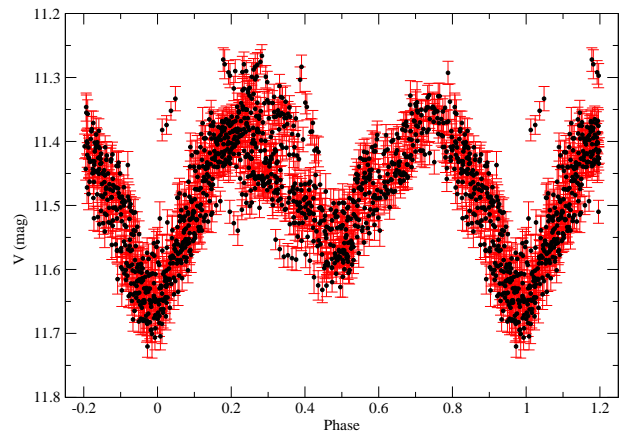}
\vspace{4mm}
      \caption{Complete INTEGRAL/OMC light curve folded with the
      period of $P = 0.4223576 \pm 0.0000009$\,d and reference epoch
      $T_0 = 2\,453\,080.0249$ (time of inferior conjunction of the
      secondary).\label{IOMC:totalomc_folded} }
\end{center}
\end{figure*}

At each INTEGRAL pointing the OMC collects a set of images with
several different integration times, typically in the range of 10\,s
to 200\,s (currently 10, 50 and 200\,s), for the purpose of optimizing
the dynamic range and also to minimize noise and cosmic-ray
effects. For the analysis of TYC~2675-663-1 we have used only the
images with exposures of 100\,s and 200\,s, in order to secure an
adequate signal-to-noise ratio for the observations. The brightness
measurements used here were obtained with the {\it Off-line Scientific
Analysis software}\footnote{Available in the web:
http://isdc.unige.ch/?Soft+download}. We used a photometric aperture
of $5\times 5$ pixels, since contamination by nearby sources was not
significant (see Figure~\ref{field}). The complete folded light curve
is shown in Figure~\ref{IOMC:totalomc_folded}.

\begin{table*}
{\scriptsize
\begin{center}
      \caption{Log of the observations used in this work.}
\label{IOMC:tobserv}
\vspace{2mm}
  \begin{tabular}{ccc}
    \hline \hline
\multicolumn{1}{c}{Time interval} &
\multicolumn{1}{c}{Type of observation} &
\multicolumn{1}{c}{Telescope}
\\
\hline
     16 Nov. 2002 --- 1 May 2005    & Photometry & INTEGRAL/OMC \\
     21 Jul. 2004 ---  1 Sep. 2004   & Photometry & 0.5\,m CAB \\
     19 Aug. 2006 --- 24 Aug. 2006   & Photometry & 1.52\,m OAN \\
     23 Sep. 2004 ---  5 Nov. 2006   & High-resolution spectroscopy & 1.5\,m Whipple Observatory \\
            25 Jul. 2007            & Low-resolution spectroscopy & 3.5\,m TWIN \\
\hline
\end{tabular}

\end{center}
}
\end{table*}

\subsubsection{Photometry from the {\it Centro de Astrobiolog\'ia}
(CAB) 0.5\,m telescope} \label{cab_phot}

Ground-based photometric observations were collected with the 0.5\,m
Giordano Bruno robotic telescope of the Centro de Astrobiolog\'{\i}a
(hereafter CAB), located at Calar Alto, Spain. This telescope is
equipped with a Finger Lakes Instrumentation IMG1024S
1024${\times}$1024 back-illuminated CCD, and Johnson $B$ and $V$
filters. The camera has an image scale of 0.97 arcsec\,pixel$^{-1}$,
which results in a field of view of $16\arcmin \times 16\arcmin$.  The
exposure times were 30\,s and 20\,s for the $B$ and $V$ filters,
respectively.  All images were corrected for bias current and flat
field. We derived differential photometry with respect to three
comparison stars in the $V$ filter (USNO B1 1224-0557970,
1224-0558767, and 1223-0560307), and an additional fourth comparison
star in $B$ (USNO B1 1224-0558836). This allowed us to obtain
differential photometry for the source with errors less than
$0.02$\,mag for both $B$ and $V$.

\subsubsection{Photometry from the {\it Observatorio Astron\'omico
Nacional} (OAN) 1.52\,m telescope}\label{oan_phot}

In order to obtain better photometric precision we carried out
additional observations with the 1.52\,m telescope at the Observatorio
Astron\'omico Nacional, also located at Calar Alto (hereafter
OAN). This telescope was equipped with a Photometrics Series 200
back-illuminated CCD, and Johnson $B$, $V$, and $I$ filters. The
camera has a scale of 0.4 arcsec\,pixel$^{-1}$, with a field of view
of $6.9\arcmin \times 6.9\arcmin$. The exposure times were 90, 60 and
30\,s for the $B$, $V$, and $I$ filters, respectively. We applied
standard bias and flat field corrections. Differential photometry was
derived with respect to three comparison stars, selected from objects
in the field with $B\!-\!V$ indices similar to the target. These stars
were USNO-B1 1225-0549546, 1225-0549756, and 1225-0549576, with
magnitudes of $B = 14.1, 15.5, 19.3$, $V = 13.45, 14.65, 18.0$, and $I
= 12.7, 13.7, 16.8$, respectively. We used a photometric aperture radius
of $7''$ for the target. The closest star to the target is located at 
$11''$ (USNO B1 1225-0549559) and is fainter, thus its contribution to the 
flux of the target is expected to be negligible.

\subsection{Spectroscopy} \label{spectroscopy}

\subsubsection{High-resolution spectroscopy} \label{highres}

TYC~2675-663-1 was observed for this project with the CfA Digital
Speedometer \citep{latham92} on the 1.5\,m Tillinghast reflector at
the Fred L.\ Whipple Observatory on Mount Hopkins, Arizona (USA). This
echelle spectrograph coupled with its intensified photon-counting
Reticon detector delivers a single echelle order 45~\AA\ wide centred
at a wavelength near 5187~\AA, and a resolving power of $R \approx
35,\!000$. The main features in this spectral window are the lines of
the Mg I b triplet. A total of 18 spectra were gathered between 23 Sep
2004 and 5 Nov 2006, with signal-to-noise ratios ranging from 14 to 30
per resolution element of 8.5 km~s$^{-1}$. ThAr exposures were taken
before and after each stellar exposure for wavelength calibration. The
velocity zero-point was monitored by means of sky exposures taken at
dusk and dawn.

\subsubsection{Low-resolution spectroscopy}

For a more detailed spectral coverage and to better study the spectral
properties of the object, a 3-hour sequence of observations of
TYC~2675-663-1 was gathered also with the 3.5\,m telescope at Calar
Alto using the double-beam spectrograph TWIN, on 25 Jul 2007 from
22:02:06 (UTC) to 26 Jul 01:02:24 (UTC). The wavelength coverage is
$\sim$4400 {\AA} to 5500 {\AA} and $\sim$6000 {\AA} to 7100
{\AA} for the blue and red channels, respectively. The spectral
resolution of these observations is 0.54\,{\AA}~pix$^{-1}$, or
$\sim$25 and 32\,km~s$^{-1}$ in the blue and red channels,
respectively. The exposure time for each spectrum was 600\,s
(resolution of 50 phase bins per orbital cycle). The median seeing was
about 2$\arcsec$. Dome flat fields and bias images were taken at the
beginning of the night, and comparison HeAr lamp spectra were taken
regularly for wavelength calibration.

\section{Data analysis} \label{analysis}

\subsection{Ephemeris} \label{IOMC:ephemeris}

An initial value of the period of TYC~2675-663-1 was derived using a
method based on the Phase Dispersion Minimization algorithm of
Stellingwerf (1978). In order to refine this period, we determined
individual times of minimum from all our photometry data
(INTEGRAL/OMC, CAB 0.5\,m, and OAN 1.52\,m) by fitting each eclipse
with a fourth degree polynomial. A total of 59 timings were obtained
(21 primary eclipses and 38 secondary eclipses), which are listed in
Table~\ref{IOMC:tephemeris}. These were then used to establish the
final period and reference epoch by solving for a linear ephemeris
using standard weighted least-squares techniques. Primary and
secondary minima were adjusted simultaneously, and the orbit was
assumed to be circular. Given the asymmetry of the eclipses, realistic
uncertainties for the individual timings are a bit difficult to
determine. Instead, we assigned reasonable initial uncertainties by
telescope, and then adjusted them by iterations so as to achieve
reduced $\chi^2$ values near unity separately for the minima from each
data set. In this way we established realistic timing errors of 0.011
days, 0.012 days, and 0.006 days for the INTEGRAL/OMC and the 0.5\,m
and 1.52\,m telescopes, respectively. The resulting period and epoch
are given by
\begin{equation}
     \begin{centering}
      P = 0.4223576 \pm 0.0000009\,{\rm days}
     \end{centering}
\end{equation}
\begin{equation}
     \begin{centering}
      T_0 = 2\,453\,080.0249 \pm 0.0014\,({\rm HJD})~,
     \end{centering}
\end{equation}
where the reference epoch $T_0$ was chosen to be close to the mean
value of all the timings. We detect no sign of period changes over the
3.7-year interval, and we adopt this ephemeris for the remainder of
the paper.

\begin{table}
{\scriptsize
\begin{center}
\caption{Measured times of mid-eclipse for TYC~2675-663-1, listed with
their errors ($\sigma$), cycle number ($E$), and $O\!-\!C$ residuals
from the adopted ephemeris.}
\label{IOMC:tephemeris}
\vspace{2mm}
  \begin{tabular}{cccccc}
    \hline \hline
\multicolumn{1}{c}{HJD$-2\,400\,000$} &
\multicolumn{1}{c}{${\sigma}$} &
\multicolumn{1}{c}{$O\!-\!C$} &
\multicolumn{1}{c}{E} &
\multicolumn{1}{c}{Year} &
\multicolumn{1}{c}{Telescope}
\\
\hline
   52595.58979 &   0.0110 &  $+$0.00910 &   $-$1147.0 &   2002.8764 &  OMC  \\
   52595.76130 &   0.0110 &  $-$0.03057 &   $-$1146.5 &   2002.8768 &  OMC  \\
   52595.99658 &   0.0110 &  $-$0.00647 &   $-$1146.0 &   2002.8775 &  OMC  \\
   52596.20278 &   0.0110 &  $-$0.01145 &   $-$1145.5 &   2002.8780 &  OMC  \\
   52596.43136 &   0.0110 &  $+$0.00596 &   $-$1145.0 &   2002.8787 &  OMC  \\
   52612.48048 &   0.0110 &  $+$0.00549 &   $-$1107.0 &   2002.9226 &  OMC  \\
   52617.96213 &   0.0110 &  $-$0.00351 &   $-$1094.0 &   2002.9376 &  OMC  \\
   52619.03278 &   0.0110 &  $+$0.01124 &   $-$1091.5 &   2002.9405 &  OMC  \\
   52620.29992 &   0.0110 &  $+$0.01131 &   $-$1088.5 &   2002.9440 &  OMC  \\
   52620.92656 &   0.0110 &  $+$0.00441 &   $-$1087.0 &   2002.9457 &  OMC  \\
   52621.76625 &   0.0110 &  $-$0.00061 &   $-$1085.0 &   2002.9480 &  OMC  \\
   52621.97140 &   0.0110 &  $-$0.00664 &   $-$1084.5 &   2002.9486 &  OMC  \\
   52622.18812 &   0.0110 &  $-$0.00110 &   $-$1084.0 &   2002.9492 &  OMC  \\
   52622.81180 &   0.0110 &  $-$0.01095 &   $-$1082.5 &   2002.9509 &  OMC  \\
   52623.03091 &   0.0110 &  $-$0.00302 &   $-$1082.0 &   2002.9515 &  OMC  \\
   52623.46205 &   0.0110 &  $+$0.00576 &   $-$1081.0 &   2002.9527 &  OMC  \\
   52624.30990 &   0.0110 &  $+$0.00889 &   $-$1079.0 &   2002.9550 &  OMC  \\
   52624.52707 &   0.0110 &  $+$0.01489 &   $-$1078.5 &   2002.9556 &  OMC  \\
   52624.71628 &   0.0110 &  $-$0.00708 &   $-$1078.0 &   2002.9561 &  OMC  \\
   52624.92494 &   0.0110 &  $-$0.00960 &   $-$1077.5 &   2002.9567 &  OMC  \\
   52737.71162 &   0.0110 &  $+$0.00760 &    $-$810.5 &   2003.2655 &  OMC  \\
   52746.76104 &   0.0110 &  $-$0.02367 &    $-$789.0 &   2003.2902 &  OMC  \\
   52797.69304 &   0.0110 &  $+$0.01424 &    $-$668.5 &   2003.4297 &  OMC  \\
   52798.32169 &   0.0110 &  $+$0.00935 &    $-$667.0 &   2003.4314 &  OMC  \\
   52798.73015 &   0.0110 &  $-$0.00455 &    $-$666.0 &   2003.4325 &  OMC  \\
   52798.95242 &   0.0110 &  $+$0.00655 &    $-$665.5 &   2003.4331 &  OMC  \\
   52799.16068 &   0.0110 &  $+$0.00363 &    $-$665.0 &   2003.4337 &  OMC  \\
   52800.62850 &   0.0110 &  $-$0.00681 &    $-$661.5 &   2003.4377 &  OMC  \\
   52800.86694 &   0.0110 &  $+$0.02046 &    $-$661.0 &   2003.4384 &  OMC  \\
   52801.26446 &   0.0110 &  $-$0.00438 &    $-$660.0 &   2003.4395 &  OMC  \\
   52806.33100 &   0.0110 &  $-$0.00613 &    $-$648.0 &   2003.4533 &  OMC  \\
   53208.40495 &   0.0124 &  $-$0.01662 &     304.0 &   2004.5542 &  0.5m  \\
   53209.46826 &   0.0124 &  $-$0.00920 &     306.5 &   2004.5571 &  0.5m  \\
   53220.45425 &   0.0124 &  $-$0.00451 &     332.5 &   2004.5871 &  0.5m  \\
   53224.49223 &   0.0124 &  $+$0.02107 &     342.0 &   2004.5982 &  0.5m  \\
   53231.42901 &   0.0124 &  $-$0.01105 &     358.5 &   2004.6172 &  0.5m  \\
   53231.65903 &   0.0124 &  $+$0.00779 &     359.0 &   2004.6178 &  0.5m  \\
   53233.53572 &   0.0124 &  $-$0.01613 &     363.5 &   2004.6230 &  0.5m  \\
   53235.44962 &   0.0124 &  $-$0.00284 &     368.0 &   2004.6282 &  0.5m  \\
   53250.43605 &   0.0124 &  $-$0.01010 &     403.5 &   2004.6692 &  0.5m  \\
   53324.15075 &   0.0110 &  $+$0.00320 &     578.0 &   2004.8710 &  OMC  \\
   53324.59421 &   0.0110 &  $+$0.02430 &     579.0 &   2004.8723 &  OMC  \\
   53330.06528 &   0.0110 &  $+$0.00472 &     592.0 &   2004.8872 &  OMC  \\
   53330.49275 &   0.0110 &  $+$0.00983 &     593.0 &   2004.8884 &  OMC  \\
   53336.39682 &   0.0110 &  $+$0.00090 &     607.0 &   2004.9046 &  OMC  \\
   53336.82269 &   0.0110 &  $+$0.00441 &     608.0 &   2004.9057 &  OMC  \\
   53342.29115 &   0.0110 &  $-$0.01778 &     621.0 &   2004.9207 &  OMC  \\
   53342.74087 &   0.0110 &  $+$0.00958 &     622.0 &   2004.9219 &  OMC  \\
   53357.09190 &   0.0110 &  $+$0.00046 &     656.0 &   2004.9612 &  OMC  \\
\hline
\end{tabular}
\end{center}
}
\end{table}

\begin{table}
{\scriptsize
\begin{center}
\addtocounter{table}{-1}
\caption[(Continued.)]{(Continued.)}
\vspace{2mm}
  \begin{tabular}{cccccc}
    \hline \hline
\multicolumn{1}{c}{HJD-2\,400\,000} &
\multicolumn{1}{c}{${\sigma}$} &
\multicolumn{1}{c}{$O\!-\!C$} &
\multicolumn{1}{c}{E} &
\multicolumn{1}{c}{Year} &
\multicolumn{1}{c}{Tel}
\\
\hline
   53966.55612 &   0.0065 &  $+$0.00265 &    2099.0 &   2006.6299 &  1.52m  \\
   53967.39777 &   0.0065 &  $-$0.00041 &    2101.0 &   2006.6322 &  1.52m  \\
   53967.61149 &   0.0065 &  $+$0.00213 &    2101.5 &   2006.6327 &  1.52m  \\
   53968.44541 &   0.0065 &  $-$0.00866 &    2103.5 &   2006.6350 &  1.52m  \\
   53969.51250 &   0.0065 &  $+$0.00253 &    2106.0 &   2006.6380 &  1.52m  \\
   53970.36807 &   0.0065 &  $+$0.01339 &    2108.0 &   2006.6403 &  1.52m  \\
   53970.56628 &   0.0065 &  $+$0.00042 &    2108.5 &   2006.6408 &  1.52m  \\
   53971.40245 &   0.0065 &  $-$0.00813 &    2110.5 &   2006.6431 &  1.52m  \\
   53971.61272 &   0.0065 &  $-$0.00904 &    2111.0 &   2006.6437 &  1.52m  \\
   53972.46836 &   0.0065 &  $+$0.00189 &    2113.0 &   2006.6460 &  1.52m  \\
\hline
\end{tabular}

\end{center}
}
\end{table}

\subsection{Radial velocities and spectroscopic orbit} \label{radialv}

\begin{figure}
\begin{center}
\includegraphics[width=8cm]{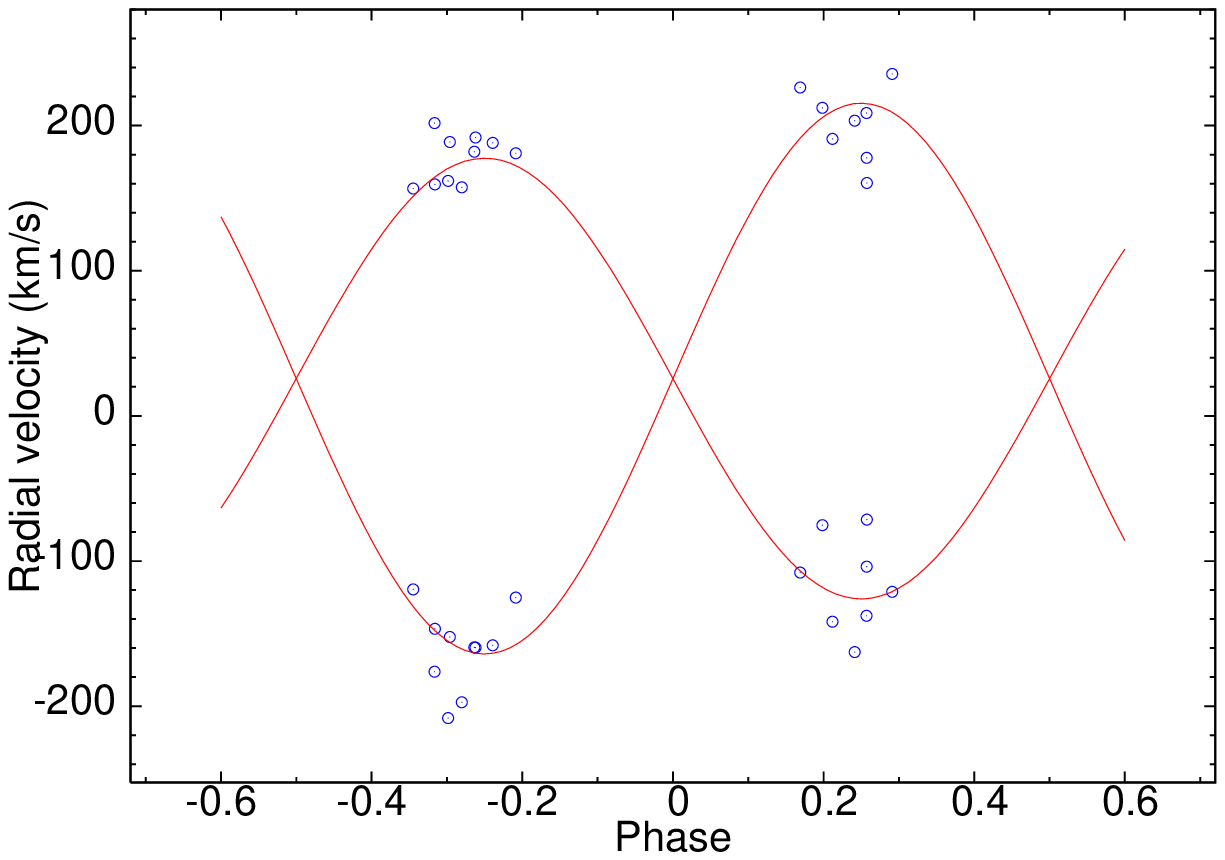}
      \caption{CfA radial velocity measurements of TYC~2675-663-1
              along with our best-fit model, folded with the period $P
              = 0.4223576 \pm 0.0000009$\,d and $T_0~{\rm (HJD)} =
              2\,453\,080.0249 \pm 0.0014$ (corresponding to phase
              0.0).
              \label{IOMC:radialvfig}  }
\end{center}
\end{figure}

Radial velocities from our high-resolution spectra were determined
using the two-dimensional cross-correlation algorithm {\tt TODCOR}
\citep{zucker94}, with templates for each star selected from a library
of synthetic spectra based on Kurucz (1992) model
atmospheres. Template parameters (effective temperature, surface
gravity, rotational velocity) were chosen to match the properties
determined for the components in the following sections, and solar
metallicity was assumed throughout. Although the typical precision for
the radial velocities of single sharp-lined stars with this
instrumentation is about 0.5 km~s$^{-1}$, in this case the performance
is significantly degraded because of the very large broadening of the
spectral lines (which we attribute to rotation) coupled with the
narrow wavelength range (single echelle order), and the double-lined
nature of the object, which combined introduce considerable line
blending. We estimate individual errors around 25\,km~s$^{-1}$ for our
measurements of the velocities of the primary and secondary
components, which are however much smaller than the large variations
we detect. These measurements are listed in
Table~\ref{IOMC:radialvtable}. A circular Keplerian orbit was fit to
these velocities holding the period and epoch of primary eclipse fixed
at the values determined earlier. The fit does not include proximity
effects, given that the observations were obtained mostly at the
quadratures where those effects are expected to be small (especially when
compared to the observational errors). We adopt these elements for
computing the absolute dimensions of the components. The resulting
orbital parameters are given in Table~\ref{IOMC:fitrvtable}, and the
observations along with our best fit are depicted in
Figure~\ref{IOMC:radialvfig}.

\begin{table}
{\scriptsize
\begin{center}
\caption{Radial velocities obtained for the primary (1) and the secondary components (2).}
\label{IOMC:radialvtable}
\vspace{2mm}
  \begin{tabular}{ccc}
    \hline \hline
\multicolumn{1}{c}{HJD$-2\,400\,000$} &
\multicolumn{1}{c}{$RV_{1}$\,(km~s$^{-1}$)} &
\multicolumn{1}{c}{$RV_{2}$\,(km~s$^{-1}$)}
\\
\hline
   53271.6872 & $+$180.87 & $-$125.13 \\
   53272.7286 & $-$103.84 & $+$177.81 \\
   53273.7860 & $+$188.05 & $-$158.08 \\
   53275.6659 & $-$141.83 & $+$190.85 \\
   53278.6415 & $-$137.75 & $+$208.62 \\
   53280.7467 & $-$162.68 & $+$203.43 \\
   53281.6124 & $-$121.20 & $+$235.58 \\
   53281.7662 & $+$156.63 & $-$119.48 \\
   53282.6452 & $+$182.05 & $-$159.53 \\
   53301.6292 & $+$159.51 & $-$146.73 \\
   53308.6291 &  $-$71.43 & $+$160.49 \\
   53684.7073 & $+$201.79 & $-$176.26 \\
   53686.6018 & $-$107.92 & $+$226.31 \\
   53690.6289 & $+$188.68 & $-$152.36 \\
   53691.6825 &  $-$75.28 & $+$212.21 \\
   53693.5844 & $+$161.95 & $-$208.17 \\
   53873.9388 & $+$157.45 & $-$197.33 \\
   54044.5790 & $+$191.78 & $-$159.58 \\
\hline
\end{tabular}

\end{center}
}
\end{table}


\begin{table}
\begin{minipage}[t]{\columnwidth}
      \caption{Spectroscopic orbital parameters of TYC~2675-663-1. }
\label{IOMC:fitrvtable}
\centering
\renewcommand{\footnoterule}{}  
\begin{tabular}{l c}
\hline\hline
 Parameter & Value \\
\hline
 $q \equiv M_{2}/M_{1}$                & $0.81 \pm 0.05$ \\
 $a \sin i$ ($R_{\odot}$)              & $2.85 \pm 0.08$ \\
 $v_{\gamma}$ (km~s$^{-1}$)            & $27 \pm 4$ \\
 $e$                                   & $0$ \footnote{Circular orbit assumed in the model.} \\
 $K_1$ (km~s$^{-1}$)                   & $153 \pm 6$ \\
 $K_2$ (km~s$^{-1}$)                   & $189 \pm 7$ \\
 $M_1 \sin^3 i$ ($M_{\odot}$)          & $0.97 \pm 0.09$  \\
 $M_2 \sin^3 i$ ($M_{\odot}$)          & $0.78 \pm 0.07$  \\
\hline
\end{tabular}
\end{minipage}
\end{table}

\begin{figure*}
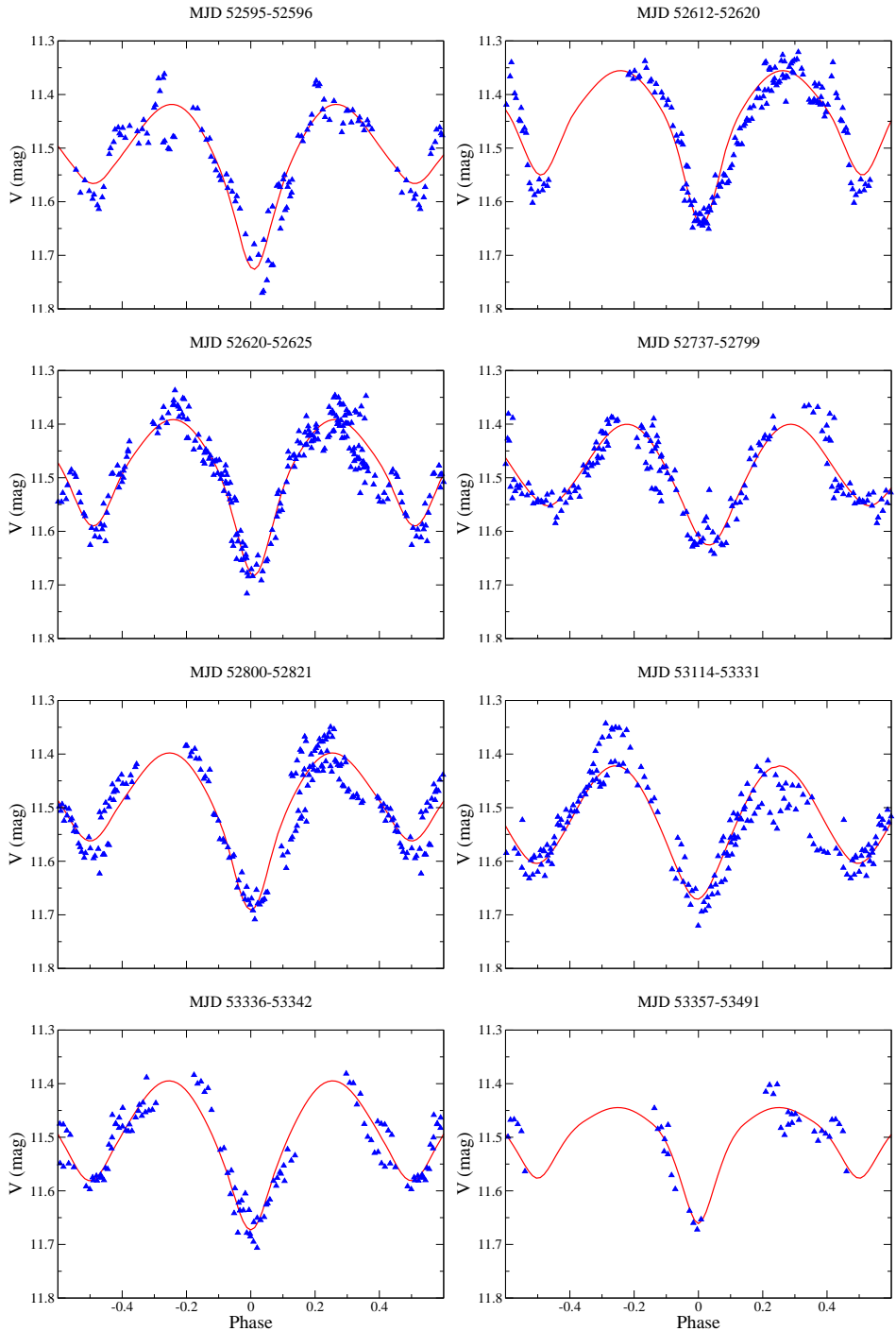

\begin{center}
\includegraphics[bb=33 54 716 577,width=6cm]{figures/f3.eps}
\hbox{   }
\includegraphics[bb=33 54 716 577,width=6cm]{figures/f4.eps}
\hbox{   }
\includegraphics[bb=33 54 716 577,width=6cm]{figures/f5.eps}
\hbox{   }
\includegraphics[bb=33 54 716 577,width=6cm]{figures/f6.eps}
\hbox{   }
\includegraphics[bb=33 54 716 577,width=6cm]{figures/f7.eps}
\hbox{   }
\includegraphics[bb=33 54 716 577,width=6cm]{figures/f8.eps}
\hbox{   }
\includegraphics[bb=33 54 716 577,width=6cm]{figures/f9.eps}
\hbox{   }
\includegraphics[bb=33 54 716 577,width=6cm]{figures/f10.eps}
\hbox{   }
      \caption{INTEGRAL/OMC light curve of TYC~2675-663-1 in the
              Johnson $V$ band, folded with the adopted
              ephemeris. Separate time intervals are shown to
              illustrate the changes in shape due to activity (from
              upper-left to lower-right). The superimposed curves
              (red) represent a fit without accounting for
              spots (see text), shown here only for reference.
              \label{IOMC:asymlc} }
\end{center}
\end{figure*}

    \begin{figure}
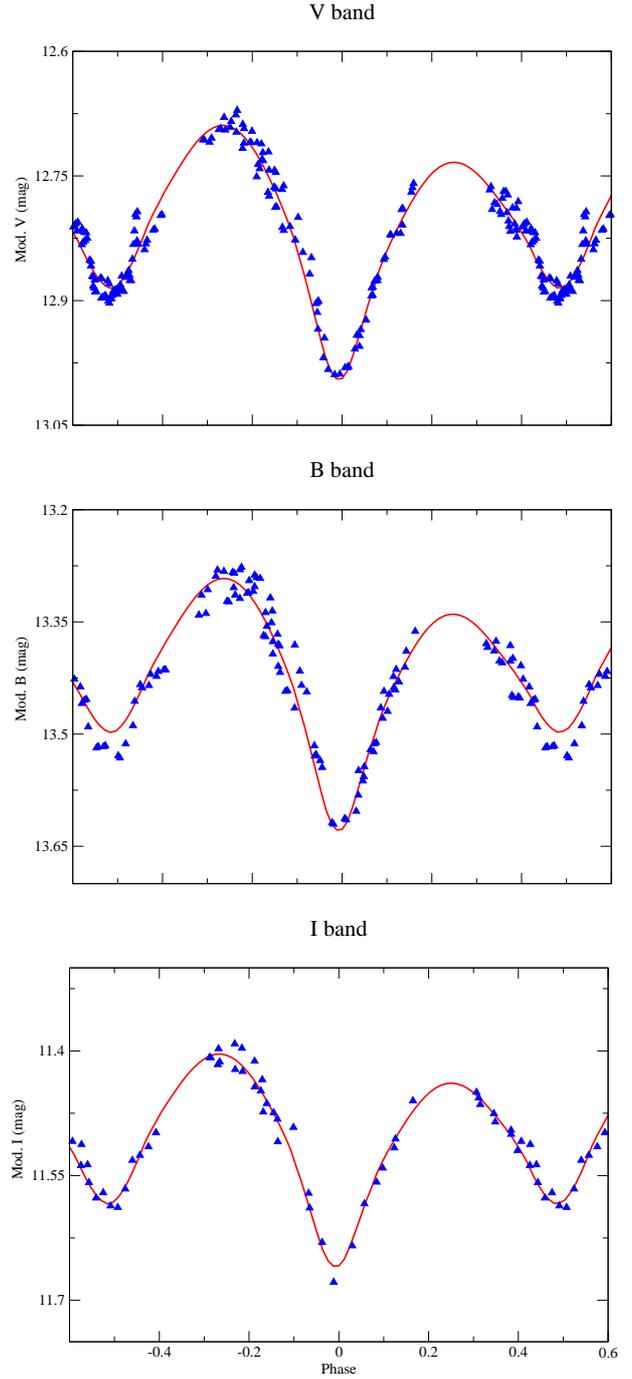

   \centering
   \includegraphics[width=8cm]{figures/f13.eps}
   \includegraphics[width=8cm]{figures/f14.eps}
   \includegraphics[width=8cm]{figures/f15.eps}
   \vspace{4mm}
      \caption{Light curves of TYC~2675-663-1 obtained with the OAN
telescope in the $V$, $B$, and $I$ bands (top to bottom), folded
according to the adopted ephemeris. This photometry is strictly
differential (magnitudes shown have an arbitrary zero point). The
curves represent our best-fit solution for an overcontact
configuration, including the effect of spots.}
         \label{cahalc}
   \end{figure}

   \begin{figure}
   \centering
   \includegraphics[bb=14 14 375 267,width=8cm,clip]{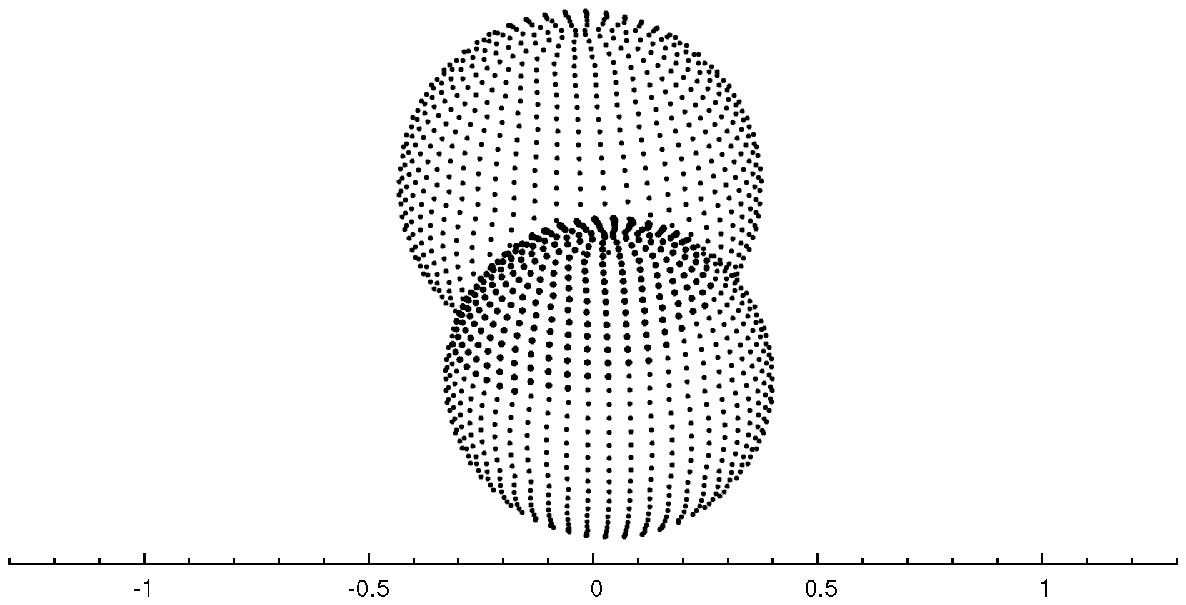}
   \includegraphics[bb=14 14 375 267,width=8cm,clip]{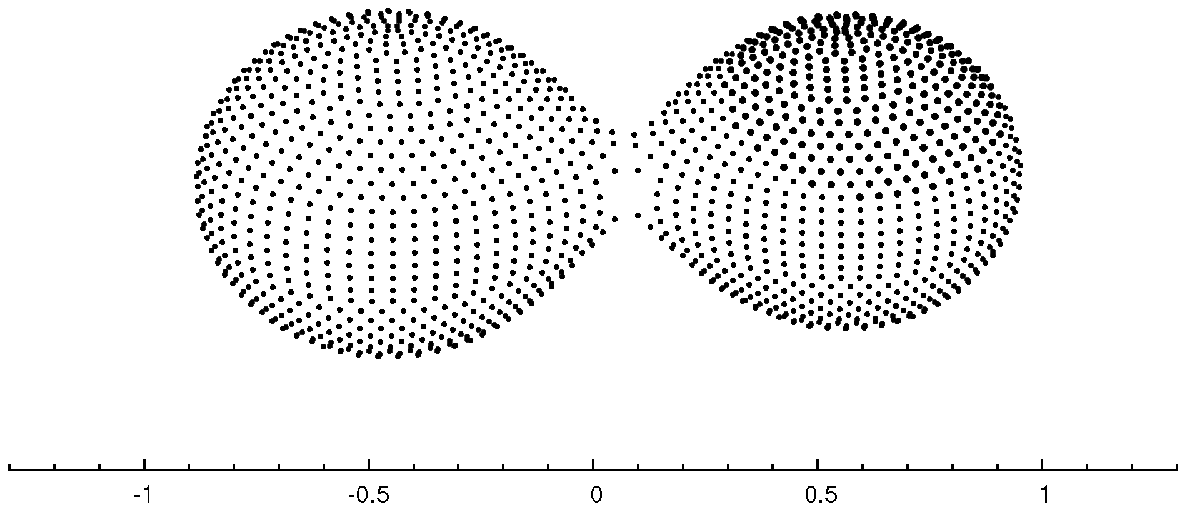}
   \includegraphics[bb=14 14 375 267,width=8cm,clip]{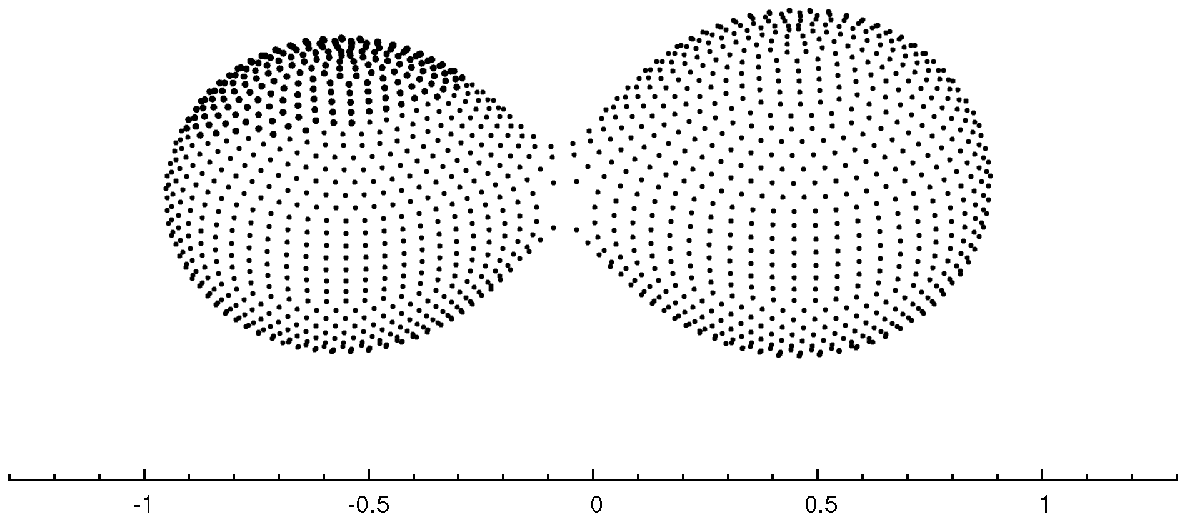}
   \caption{Configuration of the system at different phases, showing
the location of the spot on the secondary. From top to bottom, the
phases are approximately 0.00, 0.25, and 0.75. The size and separation
of the stars are rendered to scale.  }
         \label{configWD}
   \end{figure}

\begin{table}
\begin{minipage}[t]{\columnwidth}
      \caption{Light curve solutions based on the OAN photometry, for
      an overcontact and semi-detached configuration.}
\label{IOMC:fittable2}
\centering
\renewcommand{\footnoterule}{}  
\begin{tabular}{l c c }
\hline\hline
 Parameter & Overcontact (adopted)  & Semi-detached      \\
\hline
$i$ ($^{\circ}$)           &    $65.2^{+2.0}_{-3}$           &     $62.90^{+2.20}_{-0.10}$           \\
$T_{\rm eff,2}$ (K)                &        $5543^{+500}_{-24}$    &   $5724_{-25}^{+160}$         \\
$\log g_1$               &  $4.38_{-0.06}^{+0.05}$           &    $4.423_{-0.070}^{+0.015}$    \\
$\log g_2$               &  $4.36_{-0.05}^{+0.06}$           &   $4.40_{-0.07}^{+0.04}$        \\
$\Omega_1$               &    $3.37_{-0.05}^{+0.06}$         &    $3.43 \pm 0.08$            \\
$\Omega_2$               &    $\Omega_1$                     &    $3.397_{-0.030}^{+0.018}$           \\
$L_1/(L_1+L_2+L_3)_B$  &     $0.71 \pm 0.09$          &    $0.685 \pm 0.006$         \\
$L_1/(L_1+L_2+L_3)_V$  &     $0.680 \pm 0.029$        &    $0.66 \pm 0.05$              \\
$L_1/(L_1+L_2+L_3)_I$  &     $0.6272 \pm 0.0016$      &    $0.613 \pm 0.016$            \\
${\ell}_{3\, B}$ \footnote{Third-light values at phase 0.25.}   &     $0.035 \pm 0.018$           &    $0.0263 \pm 0.0027$                \\
${\ell}_{3\, V}$                 &                      $0.0399 \pm 0.0006$         &    $0.031 \pm 0.009$         \\
${\ell}_{3\, I}$                  &                      $0.0489 \pm 0.0011$         &    $0.037 \pm 0.003$        \\
${\rho}_{\rm pole\,1}$  \footnote{Relative radii in different directions, in units of the semi-major axis.} &    $0.38 \pm +0.05$        &  $0.37 \pm 0.05$     \\
${\rho}_{\rm point\,1}$ &          --                  &  $0.52 \pm 0.05$                      \\
${\rho}_{\rm side\,1}$ &   $0.40 \pm 0.05$     &  $0.39 \pm 0.05$                     \\
${\rho}_{\rm back\,1}$ &   $0.44 \pm 0.05$    &  $0.42 \pm 0.05$                     \\
$R_1$ ($R_{\odot}$)\footnote{Volume radius, defined as the radius of a sphere with the same volume as the limiting Roche lobe.} & $1.22_{-0.03}^{+0.05}$      &   $1.190 \pm 0.005$           \\
${\rho}_{\rm pole\,2}$ &     $0.35 \pm 0.05$  &   $0.34 \pm 0.05$                  \\
${\rho}_{\rm point\,2}$ &        --                    &   --                               \\
${\rho}_{\rm side\,2}$ &    $0.36 \pm 0.05$   &   $0.36 \pm 0.05$              \\
${\rho}_{\rm back\,2}$ &    $0.40 \pm 0.05$   &   $0.39 \pm 0.05$              \\
$R_2$ ($R_{\odot}$) &    $1.11^{+0.05}_{-0.03}$    &   $1.10_{-0.01}^{+0.02}$     \\
$M_{\rm bol 1}$\,(mag) &  $3.81^{+0.18}_{-0.21}$ &     $3.86 \pm 0.12$               \\
$M_{\rm bol 2}$\,(mag) &  $4.69_{-0.50}^{+0.08}$  &    $4.57_{-0.16}^{+0.04}$        \\
\multicolumn{3}{c}{Spot parameters}   \\
Colatitude$_{2}$\,(deg)      & $20$           &  $20$              \\
Longitude$_{2}$\,(deg)       & $120.0^{+8}_{-0.5}$  &  $135.2^{+1.2}_{-29}$     \\
Radius$_{2}$\,(deg)          & $57^{+47}_{-10}$    &  $74_{-23}^{+12}$                \\
$T_{\rm spot}$$_{2}$/$T_{\rm surface}$$_{2}$   &   $0.80$        &  $0.80$               \\
\hline
$\sqrt{{\Sigma}{\Delta}{\sigma}^{2} {(V)}/N}$       &  $0.001136$           &   $0.001240$  \\
$\sqrt{{\Sigma}{\Delta}{\sigma}^{2} {(B)}/N}$       &  $0.000910$           &   $0.001006$  \\
$\sqrt{{\Sigma}{\Delta}{\sigma}^{2} {(I)}/N}$        &  $0.002939$           &   $0.003108$  \\
\hline
\end{tabular}
\end{minipage}
\end{table}

\begin{table}
\begin{minipage}[t]{\columnwidth}
      \caption{Light curve solutions without accounting for spot
effects, based on the INTEGRAL/OMC and OAN photometry. These results
are for comparison purposes only, and are not our final adopted
values.}
\label{IOMC:fittable1}
\centering
\renewcommand{\footnoterule}{}  
\begin{tabular}{l c c }
\hline\hline
 Parameter & Fits to INTEGRAL/OMC & Fits to OAN \\
           & data                 & data        \\
\hline
$i$ ($^{\circ})$           &     $68_{-8}^{+11}$       &    $64.8_{-1.4}^{+1.6}$              \\
$T_{\rm eff,2}$ (K) &      $5400_{-700}^{+200}$  &    $5350_{-140}^{+230}$                 \\
$\log g_1$   &  $4.32_{-0.16}^{+0.18}$          &    $4.36 \pm 0.09$                     \\
$\log g_2$   &  $4.30_{-0.15}^{+0.19}$          &   $4.34 \pm 0.09$            \\
$\Omega_1$ &     $3.31_{-0.09}^{+0.07}$       &     $3.32_{-0.02}^{+0.03}$                    \\
$\Omega_2$ &       $\Omega_1$  &     $\Omega_1$                  \\
$L_{1}/(L_{1}+L_{2}+L_{3})_B$  &         --             &    $0.737 \pm 0.024$          \\
$L_{1}/(L_{1}+L_{2}+L_{3})_V$  &     $0.70 \pm 0.09$    &    $0.7172 \pm 0.0020$        \\
$L_{1}/(L_{1}+L_{2}+L_{3})_I$  &         --             &    $0.664 \pm 0.012$    \\
${\ell}_{3\, B}$ \footnote{Third-light values at phase 0.25.}    &         --               &  $0.031 \pm 0.006$            \\
${\ell}_{3\, V}$                                    & $0.061_{-0.015}^{+0.08}$ &  $0.0433 \pm 0.0010$               \\
${\ell}_{3\, I}$                                    &         --               &  $0.0483 \pm 0.010$     \\
${\rho}_{\rm pole\,1}$  \footnote{Relative radii in different directions, in units of the semi-major axis.} &   $0.39 \pm 0.05$  &   $0.39 \pm 0.05$       \\ 
${\rho}_{\rm point\,1}$ &     --                    &   --                                                                        \\
${\rho}_{\rm side\,1}$ &     $0.42 \pm 0.05$        &   $0.41 \pm 0.05$        \\
${\rho}_{\rm back\,1}$ &     $0.46 \pm 0.05$        &   $0.45 \pm 0.05$        \\
$R_1$ ($R_{\odot}$)\footnote{Volume radius, defined as the radius of a sphere with the same volume as the limiting Roche lobe.} &  $1.26_{-0.07}^{+0.06}$   &   $1.25 \pm 0.05$            \\ 
${\rho}_{\rm pole\,2}$ &   $0.36 \pm 0.05$          &   $0.35 \pm 0.05$      \\
${\rho}_{\rm point\,2}$ &    --                     &   --                   \\
${\rho}_{\rm side\,2}$ &   $0.38 \pm 0.05$          &   $0.37 \pm 0.05$         \\
${\rho}_{\rm back\,2}$ &   $0.42 \pm 0.05$          &   $0.42 \pm 0.05$            \\
$R_2$ ($R_{\odot}$) &     $1.16_{-0.08}^{+0.05}$     &     $1.15 \pm 0.05$      \\
$M_{\rm bol\, 1}$ (mag) &   $3.74_{-0.24}^{+0.25}$      &    $3.76 \pm 0.21$              \\
$M_{\rm bol\, 2}$ (mag) &   $4.71_{-0.25}^{+0.7}$      &    $4.71_{-0.21}^{+0.27}$               \\
\hline
\end{tabular}
\end{minipage}
\end{table}

\begin{table}
\begin{minipage}[t]{\columnwidth}
      \caption{Absolute dimensions of TYC~2675-663-1. }
\label{IOMC:dimensions}
\centering
\renewcommand{\footnoterule}{}  
\begin{tabular}{l c c }
\hline\hline
 Parameter & Primary  &  Secondary  \\
\hline
 $M$ ($M_{\odot}$)           &    $1.29^{+0.11}_{-0.06}$     &  $1.04_{-0.04}^{+0.09}$       \\
 $R$ ($R_{\odot}$)           &  $1.22_{-0.03}^{+0.05}$       &  $1.11^{+0.05}_{-0.03}$      \\
 $T_{\rm eff}$ (K)                   &   $6480 \pm 180$            &  $5543^{+500}_{-24}$    \\
 $\log g$                   &    $4.38_{-0.06}^{+0.05}$     &    $4.36_{-0.05}^{+0.06}$          \\
 $L$ ($L_{\odot}$)           &  $2.4_{-0.4}^{+0.5}$          &   $1.04_{-0.07}^{+0.6}$                          \\
 $M_{\rm bol}$ (mag)       &  $3.81^{+0.18}_{-0.21}$       &    $4.69_{-0.50}^{+0.08}$          \\
 $BC_{V}$ (mag)                  &   $-0.032_{-0.019}^{+0.009}$  &   $-0.149_{-0.005}^{+0.080}$       \\
 $M_{V}$ (mag)                   &    $3.84_{-0.22}^{+0.20}$     &    $4.84_{-0.60}^{+0.08}$  \\
\hline
 Distance (pc)                 &    \multicolumn{2}{c}{$360_{-50}^{+130}$}       \\
\hline
\end{tabular}
\end{minipage}
\end{table}

\subsection{Light curve fitting}  \label{radlcfit}

In order to characterize the system and determine its physical
properties, we have analysed the light curves taking into account the
information obtained from the spectroscopic solution just described.
Modeling of the photometry was performed with the PHOEBE package
\citep{prsa05} (version 29c), based on the Wilson-Devinney model
\citep{wilson71}. The main adjustable parameters in this model are
typically the inclination of the orbit ($i$), the
\mbox{(pseudo-)}potentials ($\Omega_1$, $\Omega_2$), the luminosity of
the primary ($L_1$) in each passband, the temperature of the secondary
($T_{\rm eff,2}$), and a phase shift. The light curves alone do not
provide any constraint on the absolute temperatures, but only on the
temperature {\it ratio} through the depth of the eclipses. Therefore,
the primary value $T_{\rm eff,1}$ was held fixed. We estimated the
temperature of the primary star from the combined $(V-K)$
colour\footnote{We adopt $V = 11.40 \pm 0.05$ and $K = 10.116 \pm
0.023$, where $V$ comes from the mean INTEGRAL/OMC standard magnitude
out of eclipse (similar to the Johnson $V$ magnitude). The
corresponding uncertainty is a conservative estimate based on the
short period variations of the light curve over timescales of a few
weeks (less affected by stellar activity, although those effects are
still present, as seen in Figure~\ref{IOMC:asymlc}). The $K$ magnitude
is taken from the 2MASS catalog \citep{cutri03}. We adopt also
reddening corrections of $E(V\!-\!K) = 0.16$ and $E(B\!-\!V) = 0.065$,
as well as $A(V) = 0.2$ \citep{fitzpatrick99}.  The temperature is
based on the $(V-K)$--$T_{\rm eff}$ relationship from
\citet{masana06}.} and estimates of the magnitude difference between
the components derived from the light curves, in an iterative way. The
result is $T_{\rm eff,1} = 6480 \pm 180$~K, where the error is a
conservative estimate based on the scatter of values found from the
colour in the iterative process of fitting the OAN light curves.

Limb-darkening coefficients were interpolated from the tables by
\citet{vanhamme93}, using the square-root law. The gravity brightening
exponents (0.3 for both stars) and bolometric albedos (0.5) were fixed
at the values appropriate for stars with convective envelopes
\citep{lucy68}.  The mass ratio $q$ and projected semimajor axis $a
\sin i$ were adopted from the spectroscopy. Because of the higher
quality of the $BVI$ photometry from the 1.52\,m OAN compared to that
from the 0.5\,m CAB, and the relatively short time span of the OAN
observations that makes them less vulnerable than the INTEGRAL/OMC
photometry to variability in the light curve, as described below, we
use only the OAN data in the following to determine the light curve
parameters. The three passbands were fitted simultaneously. The
INTEGRAL/OMC measurements were used as a consistency check.

It is obvious from the light curves (Figure~\ref{IOMC:asymlc}) that
the shape of the modulation changes continuously with phase, as in the
classical W~UMa systems, with no clear beginning or ending of the
eclipses. This strongly suggests significant deformation of the
components, and perhaps some degree of contact. Consequently, we
performed fits in both the semidetached mode (mode 4 in the
Wilson-Devinney nomenclature, with the primary filling its Roche
lobe), and also in the overcontact mode (mode 3), which is appropriate
for W~UMa-type systems and overcontact systems that are not in thermal
contact. In the former case the potential of the primary is set to the
value of its Roche lobe potential for the adopted mass ratio ($q =
0.81 \pm 0.05$): $\Omega_1 = 3.43 \pm 0.08$ (\citealt{wilson93}
\footnote{The corresponding secondary Roche lobe potential is
$\Omega_2 = 3.47 \pm 0.14$, assuming synchronous rotation.}). In the
overcontact mode the potentials are constrained to be the same, but we
do not force there to be in thermal contact, in other words, the surface
brightness of the two stars can be different even though they might be
in geometrical contact.

Initial solutions indicated a secondary temperature $T_{\rm eff,2}$ in
the range of 5100--5700~K, but also gave rather low values of the
inclination angle ($\approx 57$ deg) as a result of the relatively
shallow eclipses, leading to absolute masses of $M_1 \approx
1.62~M_{\odot}$ and $M_2 \approx 1.31~M_{\odot}$. Such large masses,
which are typical of late A- and mid F-type stars, respectively, would
imply temperatures that are considerably higher than we estimate based
on the colours (see above). Experiments in which we included third light
($\ell_3$) in our solutions revealed that it is statistically
different from zero ($> 3\sigma$), providing a plausible explanation.  
We obtained $\ell_3$ values of 2--3\% in $B$, 3--4\% in $V$, and 4--5\% in $I$,
which suggests a red object contaminating the photometry. These
solutions yielded considerably larger inclinations, as expected, by
about 5--9$^{\circ}$. We do not see any clear evidence of a third star
in our CfA or TWIN spectra, although the spectroscopic material is
inadequate for detecting such faint signatures so it does not rule
them out. In the following we have chosen to include third light in
our light curve modelling, on the basis that it is statistically
significant and provides for a more consistent overall solution.

There are abundant signs of chromospheric activity in TYC~2675-663-1,
manifested in the form of an occasional O'Connell effect, irregular
shapes of the minima, and occasional flares (see Figure
\ref{IOMC:asymlc}).  There is also significant variability in the
shape of the light curves, on typical timescales of a few weeks. The
O'Connell effect is apparent in the different light levels at the
quadratures ($\Delta V$, $\Delta B \approx 0.1$ mag), and its reality
is demonstrated by the fact that it is seen in the light curves from
the 0.5\,m CAB telescope (not shown) during one of the same time
intervals covered by the INTEGRAL/OMC photometry (Figure
\ref{IOMC:asymlc}), thus ruling out instrumental errors.  This
strongly suggests spottedness in one or both stars, which must be
considered in order to avoid biases in the geometric elements. PHOEBE
allows the spot effects to be modelled, by assuming one or more
uniform-temperature circular features parametrized by four additional
adjustable variables: the longitude, colatitude, angular size, and
temperature contrast factor ($T_{\rm spot}/T_{\rm star}$). However,
because of well-known degeneracies in fitting for spot parameters from
light curves \citep[see, e.g.,][]{Eker:96, Eker:99}, and the limited
quality of our observations, it is not possible to discern precisely
and unambiguously where the spots are in this system. For this work we
have arbitrarily chosen to place a single spot on the secondary star,
and we have chosen its location to be near the pole (colatitude $b =
20^{\circ}$), on the grounds that similar locations are often seen in
other active and rapidly-rotating stars studied, e.g., with Doppler
imaging techniques (see, e.g. \citealt{strassmeier09} and references
therein). We have also fixed the temperature contrast factor
to $T_{\rm spot}/T_{\rm star} = 0.80$, a value similar to that used in
other studies. Experiments with a spot location near the equator
produced fits of about the same quality, and also gave similar values
for all geometric and radiative quantities, within the errors. The
fits converged to a large spot covering $\sim$20\% of the surface of
the secondary star.

Solutions with third light and spots using the OAN photometry were
carried out both in the semi-detached mode and in the overcontact
mode. The fits consistently indicated contact between the components
and a slight preference for the overcontact configuration, but gave
otherwise very similar results for all parameters, the differences
being well within the errors. As seen in Table \ref{IOMC:fittable2}
the potentials of the two components in the semi-detached solution
appear nearly identical (within the errors), and equal to those from
the overcontact solution.  This fact agrees with the definition of the
overcontact mode, in which the components have the same potentials.
Therefore, we adopt in the following the overcontact results. These
solutions are presented in Table \ref{IOMC:fittable2}, and are shown
in Figure~\ref{cahalc} together with the OAN photometry. The values of
the parameters correspond to the results from fitting simultaneously
the $B$, $V$, and $I$ light curves, and the errors represent the
dispersion of the solution with respect those from fits of the
monochromatic light curves. A graphical representation of the
configuration of the system and the location of the spot is shown in
Figure~\ref{configWD}.

As a consistency check, we produced an additional fit using the
INTEGRAL/OMC data in the adopted overcontact mode. However, because of
the longer time coverage of these data compared to the OAN light
curves, and the significant spot variability on relatively short
timescales, we did not consider spots in this solution (i.e., this is
not an optimized fit, and not our adopted fit). For the purpose of
comparison, we fitted the same spotless model to the OAN light curves,
and the results are listed in Table~\ref{IOMC:fittable1}. There is
very good agreement between the two data sets, despite the simplified
modelling. This solution is over-plotted with the INTEGRAL/OMC data in
Figure~\ref{IOMC:asymlc}. A comparison for the OAN data between the
spotted model in Table~\ref{IOMC:fittable2} and the unspotted model in
Table~\ref{IOMC:fittable1} shows that the differences are rather
small, and thus that the absolute dimensions of TYC~2675-663-1 are
unlikely to be much affected by the uncertainties in the spot
modelling.

\subsection{Absolute dimensions}

The combination of the light curve and spectroscopic parameters leads
to the physical properties of the system presented in
Table~\ref{IOMC:dimensions}.  The temperature difference between the
stars is $\sim$940~K, and the individual values correspond to spectral
types of approximately F5 and G7 for the primary and secondary,
respectively.  Using these individual temperatures, we infer
bolometric corrections $BC_V$ of $-0.032^{+0.009}_{-0.019}$ and
$-0.149^{+0.080}_{-0.005}$ for the primary and secondary stars,
respectively, based on the tabulations by \citet{masana06}.  The
errors are propagated from the temperature uncertainties. With the
bolometric magnitudes that follow from the temperatures and radii, we
derive absolute visual magnitudes $M_{V}$ of $3.84^{+0.20}_{-0.22}$
and $4.84_{-0.60}^{+0.08}$.  The magnitude difference is in good
agreement with the value inferred from the flux ratio between the
components from the fits (i.e.,  $M_{V\,2} - M_{V\,1} = 2.5
\log(L_{1}/L_{2}) = 0.96^{+0.16}_{-0.15}$; see Table
\ref{IOMC:fittable2}).

The apparent visual magnitudes of the primary and secondary follow
from the out-of-eclipse brightness and the light ratio, and are $11.80
\pm 0.08$ and $12.80 \pm 0.08$, respectively. With a visual extinction
estimate of $A(V) = 0.2$ from \citet{drimmel03}, and the absolute
magnitudes given above, we infer a distance to the binary system of $D
= 360^{+130}_{-50}$ pc.

\subsection{X-ray emission}  \label{distance}

The sky position of the optical source TYC~2675-663-1 is 14\arcsec\
far from the ROSAT source 1RXS~J200912.0+323344, which has a
1$\sigma$ positional uncertainty of 8\arcsec\ \citep{voges99}. This
implies a formal coincidence likelihood between the optical and the
X-ray locations of $<13\%$. The reality of this X-ray source is
confirmed by its detection by the {\it XMM-Newton} satellite (under
the designation XMMSL1~J200910.0+323358) at similar coordinates as the
ROSAT source \citep{freyberg06} and with a similar positional
uncertainty\footnote{The quoted positions in the slew catalog have a 1
sigma error of 8$\arcsec$, which for point sources is dominated by the
accuracy of the attitude reconstruction during the slews. The
statistical position error quoted in the catalog has a mean of
$\sim$4$\arcsec$ (1 image pixel) for non-extended sources.}. The count
rate and hardness ratio of the source measured by ROSAT are $0.123 \pm
0.016$~counts~s$^{-1}$ and $0.82 \pm 0.08$, respectively. This
translates to an integrated flux of $1.56{\times}10^{-12}$\,erg\,s$^{-1}$cm$^{-2}$
in the 0.1--2.4\,keV energy band \citep{schmitt95}. Assuming a
distance to the source of $D=360_{-50}^{+130}$\,pc then gives an
intrinsic luminosity of $L_{X} =
(2.4_{-0.6}^{+2.1}){\times}10^{31}$\,erg\,s$^{-1}$. Considering the
bolometric luminosities of the components of $L_{\rm bol} =
2.4^{+0.5}_{-0.4}$\,$L_{\odot}$ and $1.04^{+0.6}_{-0.07}$\,$L_{\odot}$
for the primary and the secondary, we derive an X-ray to total
luminosity ratio of $\log L_X/L_{\rm bol} = -2.74^{+0.24}_{-0.40}$,
which is typical of binary systems with rotationally induced activity
(see, e.g., \citealt{messina03}).

\subsection{Analysis of spectroscopic features} \label{highresspec}

\begin{figure}
   \centering
   \includegraphics[width=8cm]{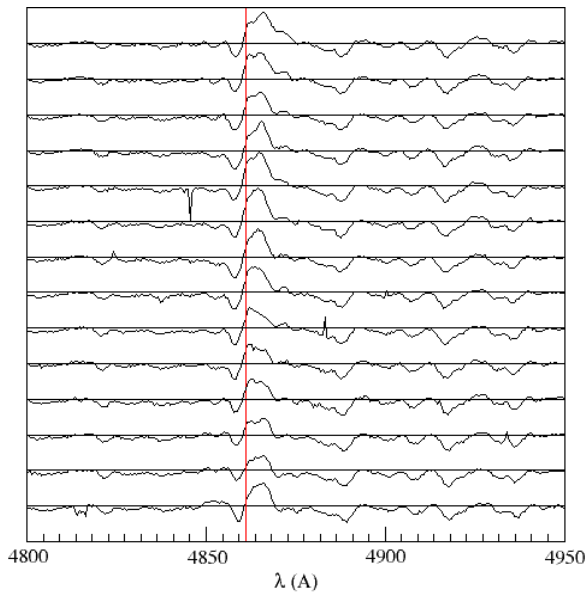}
      \caption{Blue spectra of TYC~2675-663-1 obtained with the TWIN
              spectrograph (4800--4950\,{\AA}) showing the
              H$_\beta$ emission complex centred at $\lambda_0 =
              4861$\,{\AA}. Orbital phases for each spectrum are as
              indicated in Figure \ref{redspe}.}
         \label{bluespe}
   \end{figure}

\begin{figure}
   \centering
   \includegraphics[bb=73 2 569 602,width=8cm,clip]{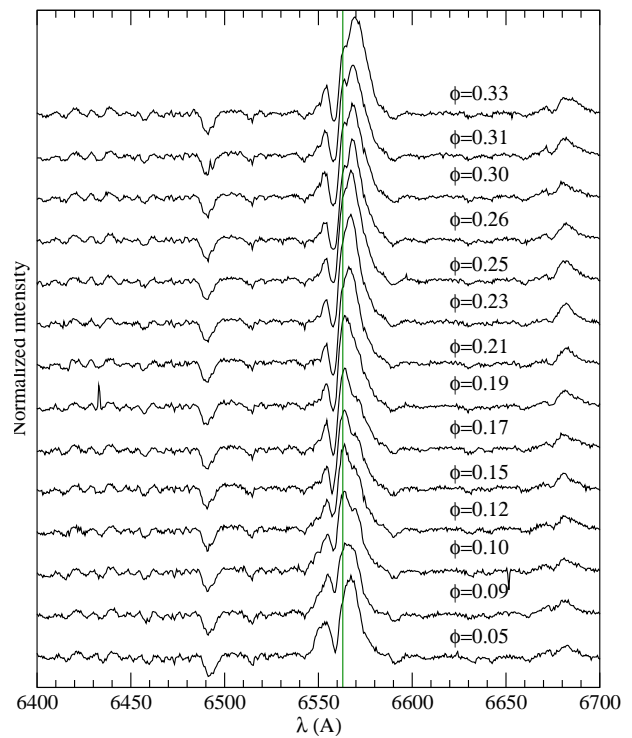}
      \caption{Red spectra of TYC~2675-663-1 obtained with the TWIN
              spectrograph (6400--6700\,{\AA}) showing the
              prominent H$_\alpha$ and He~{\sc i} emission features,
              centred at $\lambda_0 = 6563$\,{\AA} and $\lambda_0 =
              6680$\,{\AA}, respectively. In this figure the
              wavelength shift due to the orbital motion of the
              components is approximately 4\,{\AA}. Orbital phases are
              labelled.}
         \label{redspe}
   \end{figure}

Figures~\ref{bluespe} and \ref{redspe} display the sequence of 14
normalized low-resolution spectra from the TWIN spectrograph in the
range from 4800 to 4950 \AA\ and from 6400 to 6700 \AA, showing
the most interesting features. Among the most prominent are the strong
Balmer lines in emission (H$_\alpha$ and H$_\beta$ at $\lambda =$
6563 \AA\ and 4861\,\AA, respectively) and the He~{\sc i} line
centred at 6680\,\AA. The H$_\alpha$ emission is complex and is
discussed below.

\begin{figure}
   \centering
   \includegraphics[width=8cm]{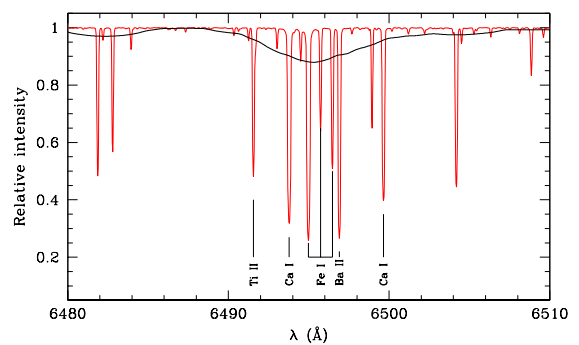}
      \caption{Synthetic spectrum corresponding to a star
              similar to the primary of TYC~2675-663-1, with $T_{\rm
              eff} \approx 6000$\,K, $\log g = 4.0$, [M/H] $= 0.0$,
              and no rotational broadening (red line). Superimposed is
              the same spectrum broadened with a rotation 
              profile of 144 km~s$^{-1}$ (black line), estimated
              for the primary in TYC~2675-663-1.}
         \label{syn_spe}
   \end{figure}

In the red spectra (Figure \ref{redspe}) there is a clear absorption
feature at around 6490\,{\AA}. If this has a photospheric origin,
it is likely to be a blend composed mainly of Ti~{\sc ii}, Ca~{\sc i},
Fe~{\sc i}, and Ba~{\sc ii} lines, which are the ones with the largest
oscillator strengths in this spectral range. To illustrate this, in
Figure~\ref{syn_spe} we show the synthetic spectrum of a star
similar to the primary component of TYC~2675-663-1 ($T_{\rm eff} \sim
6000$\,K, $\log g = 4.0$, $v \sin i = 0$\,km~s$^{-1}$, and an
assumed [M/H] $= 0.0$), together with the result of applying a
rotational broadening of 144\,km~s$^{-1}$, which is our estimate for
the primary star assuming spin-orbit synchronization. This washes out
the narrow lines, producing a single broad feature, as observed.
The steady character of this broad absorption suggests its origin 
in a third star, perhaps a distant companion to the binary system. This 
may perhaps explain the third light contribution inferred from the fits 
of the light curves in Section \ref{radlcfit}. Nevertheless, more complete
spectroscopic observations are needed in order to confirm the 
likely stellar origin of this absorption complex. 

\subsubsection{The H$_{\alpha}$ emission line}  \label{halpha}

\begin{figure}
   \centering
   \includegraphics[bb=71 1 569 602,width=8cm,clip]{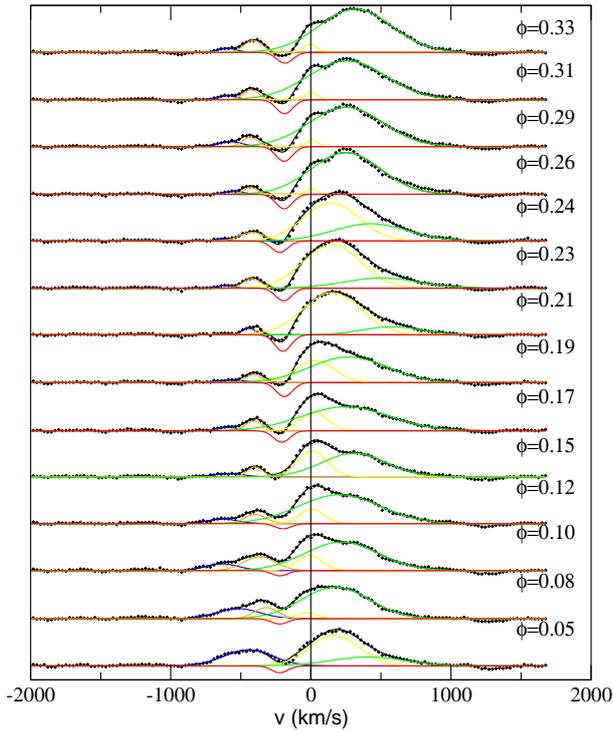}
      \caption{Enlarged view of the H$_\alpha$ complex from
      Figure~\ref{redspe}, with our 5-component fits (one absorption
      component, and 4 emission components; see text). Orbital phases
      are labelled.}
         \label{redcomp}
   \end{figure}

\begin{figure}
   \centering
   \includegraphics[bb=31 53 716 528,width=8cm,clip]{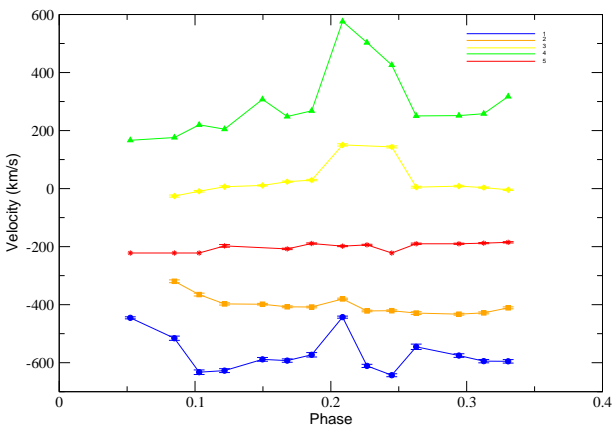}
      \caption{Radial velocity variation of each line component fitted
      to the H$_\alpha$ complex in Figure~\ref{redcomp}, as a function
      of orbital phase. The red line represents the absorption
      component, and the other four lines are in emission (see text).}
         \label{redcomp2}
   \end{figure}

Very broad H$_\alpha$ emission (${\Delta v} \approx
1200$\,km~s$^{-1}$) is detected in the spectra of this binary
system. Phenomenologically, the H$_\alpha$ complex can be described by
the presence of what appear to be two broad and asymmetric emission
lines centred at $6553$\,\AA\ and $6568$\,\AA, plus a steady 
absorption dip between them (P-Cygni profile). This emission has a
complicated structure, with several components that seem to be moving
with orbital phase. In order to better understand this complex
evolution with phase we have fitted the profile with additional
components. After trying a number of different combinations, we find
that the behaviour of the H$_\alpha$ region of the spectrum can be
adequately described with five components (see Figures~\ref{redcomp} and
\ref{redcomp2}): one in absorption and four in emission. The
absorption line (line 5) is steady, with a radial velocity of $RV
\approx -200$\,km~s$^{-1}$.

Similar outflows have been associated with stellar winds in high mass
or post-AGB stars \citep{varricat04, filliatre05}, with velocities
similar to those we find (e.g., $-105$\,km~s$^{-1}$;
\citealt{smolinski93}). However, in the case of the present binary
system TYC~2675-663-1, which is composed of two main sequence stars of
late spectral type (and likely a third star of similar or later
spectral type), the absorption is unlikely to be due to stellar
winds. Its nature is presently unclear.

In Figure \ref{redcomp2} two of the Gaussian components (lines~3 and
4) appear to follow the orbital motion of the primary star, with a
maximum velocity excursion at quadrature ($\phi = 0.25$). The
velocities at quadrature are $RV \approx 150$ km~s$^{-1}$ and $RV
\approx 575$ km~s$^{-1}$, respectively. Given the previously measured
orbital semiamplitudes of $153 \pm 6$\,km~s$^{-1}$ and $189 \pm
7$\,km~s$^{-1}$ for the primary and secondary, respectively, we
tentatively identify line~3 with the surface of the primary star.  The
emission represented by line~4 comes from a region that is moving 3--4
times faster. Elucidating the origin of these wings and the absorption
feature of the H$_\alpha$ complex will most likely require further
spectroscopic observations that are beyond the scope of the present
work, in order to obtain complete phase coverage over a full orbital
cycle.

\section{Discussion} \label{discuss}

Our analysis of the close eclipsing binary TYC~2675-663-1 has revealed
properties similar in many respects to those of the W~UMa systems,
which are characterized by having short orbital periods (0.2--0.8\,d)
and an overcontact configuration, and are composed of F--K stars
sharing a common envelope that thermalizes the stars. This leads to
near equal depths for the eclipses. TYC~2675-663-1 displays eclipses
with clearly different depths, which would imply non-thermal
equilibrium. Nevertheless, the stars have spectral types of F5 and G7
in the typical range for the W~UMa class, and appear to be in
geometrical contact despite the temperatures being different by
approximately 940\,K. The large mass ratio we derive, $q \equiv
M_2/M_1 = 0.81 \pm 0.05$, identifies the object as an H-type W~UMa
variable (``H'' for {\em high} mass ratio), a subgroup first proposed
by \citet{csizmadia04b}. In these objects the energy transfer rate
appears to be less efficient than in other types of contact
binaries. They also show excess angular momentum, which can be
understood as resulting from the first stages of the interaction
between the components \citep[see][and references therein]{li04}.
The overall properties of TYC~2675-663-1 suggest a system at or
near contact and perhaps in the early stages of evolution toward a
state of full geometrical and thermal contact (coalescence).

W~UMa systems such as the binary studied here typically show increased
chromospheric activity due to the rapid rotation of the components
\citep{applegate92}, and this is usually accompanied by soft X-ray
emission \citep{messina03}. This chromospheric activity is manifested
by the unequal brightness at quadratures, asymmetrical minima, and
erratic flares shown in the light curves, which are collectively
referred to as the O'Connell effect. TYC~2675-663-1 shows clear
evidence of each of these, including the X-ray emission. The presence
of the He\,{\sc i}~$\lambda$\,6680\,\AA\ emission line, as we see in
the spectra of TYC~2675-663-1, is often interpreted as evidence of the
interaction between the components of the binary system, in the form
of winds or streams of matter (\citealt{graham92, greeley99, takami01,
scholz06}). The H$_\alpha$ region of the spectrum is complex, with
emission and absorption components that are not yet fully understood.

Our spectroscopic and photometric observations of TYC~2675-663-1 have
provided a first picture of this close binary system in which the two
stars differ greatly in temperature, but are in a near-contact
configuration perhaps leading to coalescence. It appears to be a rare
example of this class of W~UMa objects. There is circumstantial
evidence for a third star in the system. It is hoped that this study
will be helpful in the development of theories to understand the early
behaviour and evolution of W~UMa systems.

\begin{acknowledgements}
We are grateful to the referee, A.\ Pr{\v s}a, for very helpful
comments.  MCG thanks K.\ Yakut for very useful discussions, and F.\
Vilardell for providing the table of the critical potentials. OMC has
been funded by the Spanish MCyT under grants ESP95-0389-C02-02,
ESP2002-04124-C03-01, ESP2005-07714-C03-03, and AYA2008-03467/ESP.  We
thank David Galadi and Teresa Eibe for the data taken at the CAB {\em
Giordano Bruno} telescope. This research was partially supported by
the MCyT under grant PNE2003-04352. This work is based on observations
made with INTEGRAL, an ESA science mission with instruments and
science data centre funded by ESA member states and with the
participation of Russia and the USA. We are grateful to the Calar Alto
Observatory for allocation of Director's discretionary time to this
programme.
\end{acknowledgements}


\begin{thebibliography}{}

\bibitem[Applegate (1992)]{applegate92}
Applegate, J.~H., 1992, ApJ, 385, 621
\bibitem[Caballero(2004)]{caballero04}
Caballero, M. D., 2004, Proceedings to the "5th INTEGRAL Workshop on the INTEGRAL Universe", 2004, ESA Special Publication, eds.: Schoenfelder, V., and Lichti, G. \& Winkler, C. vol. 552, p. 875 
\bibitem[Caballero-Garc\'{\i}a et~al.(2006)]{caballero06}
Caballero-Garc\'{\i}a, M.~D., Domingo, A., R\'{\i}squez, D. \& Mas-Hesse, J.~M., Proceedings to the "The X-ray Universe 2005", 2006, ed.: Wilson, A., vol. 604, p. 249
\bibitem[Csizmadia et~al.(2004)]{csizmadia04b}
Csizmadia, S., Klagyivik, P., 2004, A\&A,  426, 1001
\bibitem[Cutri et~al.(2003)]{cutri03}
Cutri, R.~M., Skrutskie, M.~F., van Dyk, S., Beichman, C.~A., Carpenter, J.~M., Chester, T., et~al., 2003, The IRSA 2MASS All-Sky Point Source Catalog, NASA/IPAC Infrared Science Archive.
\bibitem[Domingo et~al.(2003)]{domingo03}
Domingo, A., Caballero, M.~D., Figueras, F., Jordi, C., Torra, J., Mas-Hesse, J.~M., Gim{\'e}nez, A., Hudcova, V. \& Hudec, R., 2003, A\&A, 411L, 281D
\bibitem[Drimmel et~al.(2003)]{drimmel03}
Drimmel, R., Cabrera-Lavers, A. \& L{\'o}pez-Corredoira, M., 2003, A\&A, 409, 205
\bibitem[Eker(1996)]{Eker:96}
Eker, Z., 1996, ApJ, 473, 388
\bibitem[Eker(1999)]{Eker:99}
Eker, Z., 1999, ApJ, 512, 386
\bibitem[Chaty \& Filliatre(2005)]{filliatre05}
Chaty, S. \& Filliatre, P., 2005, Ap\&SS, 297, 235C
\bibitem[Fitzpatrick(1999)]{fitzpatrick99}
Fitzpatrick, E.~L., 1999, PASP, 111, 63 
\bibitem[Freyberg et~al.(2006)]{freyberg06}
Freyberg, M.~J., Altieri, B., Bermejo, D., Esquej, M.~P. et~al., 2006, Proceedings to the "The X-ray Universe 2005", ed.: Wilson, A., vol. 604, p. 913
\bibitem[Graham(1992)]{graham92}
Graham, J.~A., 1992, PASP, 104, 479G
\bibitem[Greeley et~al.(1999)]{greeley99}
Greeley, B.~W., Blair, W.~P., Long, K.~S. \& Raymond, J.~C., 1999, ApJ, 513, 491
\bibitem[H{\o}g et~al.(2000)]{hog00}
H{\o}g, E., Fabricius, C., Makarov, V.~V. et~al., 2000, A\&A, 355L, 27
\bibitem[Li et~al.(2004)]{li04}
Li, L., Han, Z. \& Zhang, F., 2004, MNRAS, 355, 1383
\bibitem[Latham(1992)]{latham92}
Latham 1992; Latham, D. W., 1992, in IAU Coll. 135, Complementary Approaches to Double and 
Multiple Star Research, ASP Conf. Ser. 32, eds. H. A. McAlister \& W. I. Harkopf (San Francisco: ASP), 110.
\bibitem[Lucy(1968)]{lucy68}
Lucy, L.~B., 1968, ApJ, 153, 877
\bibitem[Masana et~al.(2006)]{masana06}
Masana, E., Jordi, C. \& Ribas, I., 2006, A\&A, 450, 735
\bibitem[Mas-Hesse et~al.(2003)]{mas-hesse03}
Mas-Hesse, J. M., Gim\'enez, A., Culhane, J. L., et~al., 2003, A\&A, 411, L261-L268
\bibitem[Messina et~al.(2003)]{messina03}
Messina, S., Pizzolato, N., Guinan, E.~F. \& Rodon{\`o}, 2003, A\&A, 410, 671M
\bibitem[Prsa \& Zwitter(2005)]{prsa05}
Prsa, A. \& Zwitter, T., 2005, ApJ 628, 426
\bibitem[Scholz \& Jayawardhana(2006)]{scholz06}
Scholz, A. \& Jayawardhana, R., 2006, ApJ, 638, 1056
\bibitem[Schmitt et~al.(1995)]{schmitt95}
Schmitt, J.~H.~M.~M., Fleming, T.~A. \& Giampapa, M.~S., 1995, ApJ, 450, 392
\bibitem[Smolinski et~al.(1993)]{smolinski93}
Smolinski, J., Climenhaga, J.~L., Huang, Y., Jiang, S., Schmidt, M. \& Stahl, O., 1993, Space Science Reviews, 66, 231
\bibitem[Stellingwerf(1978)]{stellingwerf78} Stellingwerf, R .\ F. 1978, \apj, 224, 953
\bibitem[Strassmeier (2009)]{strassmeier09} Strassmeier, K. G. 2009, Astron. Astrophys. Rev., 17, 251 
\bibitem[Takami et~al.(2001)]{takami01}
Takami, M., Bailey, J., Gledhill, T.~M., Chrysostomou, A. \& Hough, J.~H., 2001, MNRAS, 323, 177 
\bibitem[van Hamme(1993)]{vanhamme93}
van Hamme, W., 1993, AJ, 106, 2096
\bibitem[Varricatt et~al.(2004)]{varricat04}
Varricatt, W.~P., Williams, P.~M., Ashok, N.~M. et~al., 2004, MNRAS, 351, 1307
\bibitem[Voges et~al.(1999)]{voges99}
Voges, W., Aschenbach, B., Boller, T. et~al., 1999, A\&A, 349, 389
\bibitem[Wilson \& Devinney(1971)]{wilson71} Wilson, R.~E. \& Devinney, E.~J., 1971, ApJ, 166, 605
\bibitem[Wilson(1993)]{wilson93} Documentation of Eclipsing Binary Computer Model, (University of Florida, Gainesville)
\bibitem[Zucker \& Mazeh(1994)]{zucker94}
Zucker, S. \& Mazeh, T., 1994, ApJ, 420, 806



\end{thebibliography}
\end{document}